\documentclass[
reprint,
 amsmath,amssymb,
 aps,
pra,
]{revtex4-1}

\usepackage{graphicx}
\usepackage{dcolumn}
\usepackage{bm}

\DeclareMathAlphabet{\mathpzc}{OT1}{pzc}{m}{it} 
\usepackage{braket,color,comment,amsmath}

\begin{document}

\preprint{APS/123-QED}

\title{Influence of photic perturbations on circadian rhythms}

\author{Andrea Auconi$^{1}$, Patrick Pett$^{2}$, Edda Klipp$^{1}$, and Hanspeter Herzel$^{2,}$}
\email{h.herzel@biologie.hu-berlin.de}

\affiliation{$1$ Theoretische Biophysik, Humboldt-Universit\"at zu Berlin, Germany\\
	$2$ Institute for Theoretical Biology, Charit\'e and Humboldt-Universit\"at zu Berlin, Germany}

\date{\today}

\begin{abstract}
	The circadian clock is the molecular mechanism responsible for the adaptation to daily rhythms in living organisms. Oscillations and fluctuations in environmental conditions regulate the circadian clock through signaling pathways. We study the response to continuous photic perturbations in a minimal molecular network model of the circadian clock, composed of 5 nonlinear delay differential equations with multiple feedbacks. We model the perturbation as a stationary stochastic process, and we consider the resulting irreversibility of trajectories as a key effect of the interaction. In particular we adopt a measure of mutual mapping irreversibility in the time series thermodynamics framework, and we find 12 hours harmonics.
\end{abstract}

\maketitle


The circadian clock is a network of genes whose expression levels oscillate with a period of roughly 24 hours.
Such genes interact with complex transcriptional feedback regulation \cite{becker2004modeling,klipp2016systems}. A minimal model was provided in  \cite{korenvcivc2014timing}, and it consists of delay differential equations for the five coarse-grained variables $\vec{y}\equiv (Bmal1,Per2,Cry1,Rev$-$erb\alpha,Dbp)$, each one representing more than just one gene transcript. These variables measure concentrations, so they are taken to be positive, $y_i>0~\forall i$. The collective dynamics produces deterministic self-sustained oscillations, with a limit cycle whose amplitudes and phase-differences between genes can be tuned to reproduce heterogeneity of different tissues \cite{pett2018co}. The dynamics is composed of degradation terms which are simply linear in the concentrations, and of production terms which are modeled as products of activation and repression functions of Michaelis-Menten type. While the model structure is based on biological knowledge, its parameters are optimized using experimental time-resolved quantitative data from mammalian tissues. The delays in the differential equations describe the intermediate steps required for gene interactions, like protein production, complex formation, or nuclear translocalization. Each variable $y_i$ regulates the dynamics of other variables (and of its own) with a different time delay $\tau_i$. Among the many interactions in the model, a network motif was identified as the main driving force of self-sustained oscillations \cite{pett2016feedback}, this being the repressilator loop of the three subsequent inhibitions $Per2 \dashv Rev$-$erb\alpha \dashv Cry1 \dashv Per2$.

Even though the basic period of circadian rhythms is close to 24 hours, nonlinear interactions can generate harmonics with periods of 12 hours and
8 hours \cite{westermark2013mechanism,korenvcivc2014timing}. Such harmonics have also been found experimentally \cite{hughes2009harmonics,ananthasubramaniam2018ultradian}, and
have functional relevance in metabolism \cite{cretenet2010circadian}.
Furthermore, harmonics might play a role in the adaptation to tidal
cycles with a period of 12.4 hours \cite{westermark2013mechanism}.

Photic perturbations on the mammalian circadian system are perceived in the core suprachiasmatic nucleus through induction of $Per2$ genes, this mechanism being most sensitive during the night \cite{golombek2010physiology,yan2009expression,reppert2001molecular}.
Phase-response curves \cite{granada2009phase} represent the phase lag on oscillations induced by pulse-like perturbations, as a function of the particular phase instant at which the perturbation is applied.
The phase lag is the integrated response to the pulse, and is the long-term effect of relaxation to the limit cycle from a nonequilibrium configuration. Note that in general for each choice of pulse intensity and direction (type) a different phase-response curve is obtained.

We develop a coarse-grained characterization of photic perturbations on circadian rhythms to summarize dynamical properties of the response to generic perturbations beyond the pulse scheme. These are described as (large) fluctuations leading the system constantly out-of-equilibrium, and can be formalized as stationary stochastic processes. Here we refer to photic perturbations, but the origin of noise to affect circadian rhythms in an organism can also derive from irregular feeding, activity, hormonal rhythms, and temperature \cite{abraham2018quantitative}.

The photic perturbation is an asymmetric interaction, meaning that the perturbation dynamics is not affected by the circadian genes' dynamics, and all the feedbacks are endogenous of the circadian system. This asymmetric structure is simply referred to as signal-response model. The macroscopic effect of the asymmetric interaction is the information that continuously flows from the signal $x_t$ (photic state at a time instant $t$) to the response $y_{t+\tau}$ (evolution of the circadian variables state after a generic interval $\tau$).
The conditional mutual information \cite{schreiber2000measuring,cover2012elements} or transfer entropy $I(x_t,y_{t+\tau}|y_t)$ quantifies the increase in predictive power on the response evolution $y_{t+\tau}$ that is gained upon knowledge of the signal current state $x_t$, conditional on the knowledge of the response current state $y_t$. Importantly, the transfer entropy $I(x_t,y_{t+\tau}|y_t)$ is not a measure of \textit{information flow} (or causal influence) because of its synergistic effects \cite{barrett2015exploration,auconi2017causal}, that are mirrored in the inequality $I(x_t,y_{t+\tau}|y_t)>I(x_t,y_{t+\tau})$. Quantitative definitions of information flow, synergy, and redundancy are currently under debate in the partial information decomposition framework \cite{james2016information}, and a general agreement is still missing. 

We will quantify the influence of photic perturbations on circadian rhythms studying another key aspect of the trajectories resulting from signal-response models, that is their temporal asymmetry or \textit{irreversibility}. Indeed effects are always observed after their causes, and this creates a temporal order that makes time-reversal trajectories statistically different from the original ones \cite{jarzynski2011equalities,auconi2019information}.

Information thermodynamics \cite{parrondo2015thermodynamics,seifert2012stochastic} is the study of irreversible dynamics and fluctuations in (nonequilibrium) stochastic processes. In particular, fluctuation theorems have been introduced to relate entropy production (and dissipation) with information-theoretic measures in bipartite (or multipartite) systems \cite{ito2013information,ito2016backward,horowitz2014thermodynamics,rosinberg2016continuous}. In a recent work \cite{auconi2019information} we developed a \textit{time series} formulation of information thermodynamics, and discussed fluctuation theorems on bivariate (but not necessarily bipartite) signal-response models.
In particular we introduced the mapping irreversibility $\varPhi_{\tau}^{xy}$ as a Markovian (memoryless) approximation of the time series irreversibility introduced in  \cite{roldan2012entropy}. Similarly we here define the mutual mapping irreversibility $\Theta_\tau^{xy}$ as the Markovian approximation of the mutual entropy production introduced in  \cite{diana2014mutual}, that is the mapping irreversibility of the joint process subtracted by those of the two subsystems, $\Theta_\tau^{xy}\equiv \varPhi_{\tau}^{xy}-\varPhi_{\tau}^{x}-\varPhi_{\tau}^{y}$.

\begin{figure}
	\begin{center}
		\includegraphics[scale=0.55]{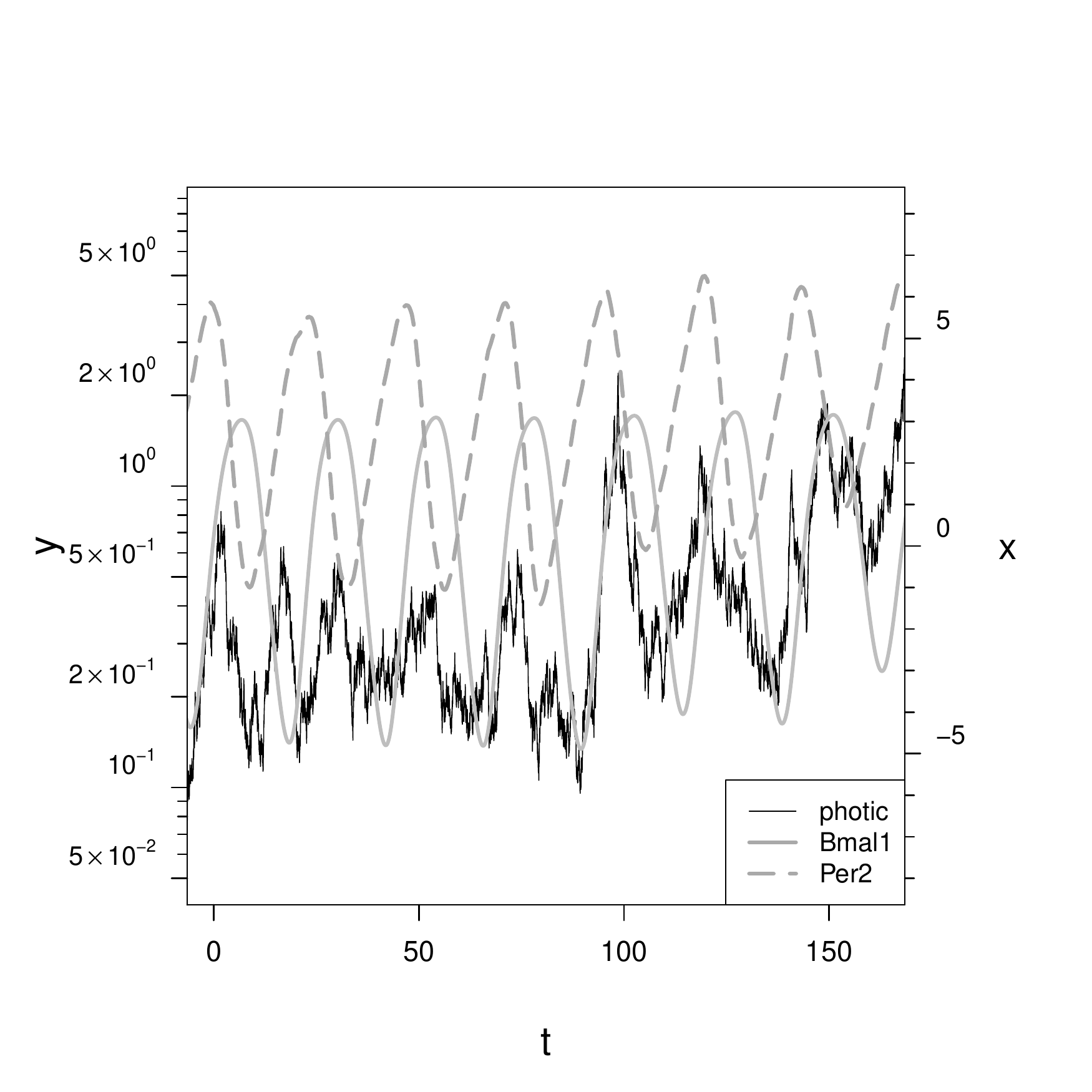}
		\caption{\label{one}Stochastic dynamics of the circadian clock genes $Per2$ and $Bmal1$ with the photic perturbation $x$ fluctuating with intensity $\gamma=0.05$ and relaxation time $t_{rel}=10$h. It is a sample of 7 days from a numerical simulation of the non-Markovian stochastic process \eqref{Circadian Model}. Time is measured in hours.}
	\end{center}
\end{figure}

We model the time-continuous photic perturbation as multiplicative noise on the production rate of $Per2$ mRNA with intensity parameter $\gamma$. As noise source we take correlated fluctuations $x$ described by an Ornstein-Uhlenbeck process \cite{uhlenbeck1930theory}, that is the simplest model of dynamical stationary fluctuations. The characteristic time of fluctuations we fix to $t_{rel}=10$h, that is compatible with the average time of environmental changes experienced by the circadian clock. The equation system for the photic perturbation $x$ and genes $\{y_i\}_{\tiny i=1,..,5}$ dynamics reads:
\begin{eqnarray}\label{Circadian Model}
\begin{cases} 
dx =-\frac{x}{t_{rel}} dt + dW \\
\frac{dy_i}{dt} = f_i(\{y_j(t-\tau_j)\}_{\small j=1,...,5}) -d_i y_i + \delta_{i2} y_i \gamma x~~
\end{cases}
\end{eqnarray}
where the $d_i$s are linear degradation coefficients, and the Kronecker delta $\delta_{i2}$ selects the photic perturbation to act only on $Per2$. $dW$ represents Brownian motion \cite{shreve2004stochastic}, which is specified by $\langle dW(t_k) dW(t_{k'}) \rangle= \delta_{k k'} dt$. The exact form of the regulating functions $f_i$ and the corresponding parameter values can be found in  \cite{korenvcivc2014timing}, where a consensus model averaging parameters of mammalian liver and adrenal gland tissues is extracted. The photic perturbation is the irregular variation to the standard periodic day/night light alternation, and is therefore modeled as a fluctuating but not oscillating process.
For a fixed correlation time $t_{rel}$, the strength of fluctuations is tuned by the parameter $\gamma$. 
The influence of standard 24 hours periodic light oscillations is considered to be already described in the deterministic model $\vec{f}$. Alternatively, the system \eqref{Circadian Model} can be considered to model perturbations to a constant darkness (DD) experiment.

\begin{figure}
	\begin{center}
		\includegraphics[scale=0.55]{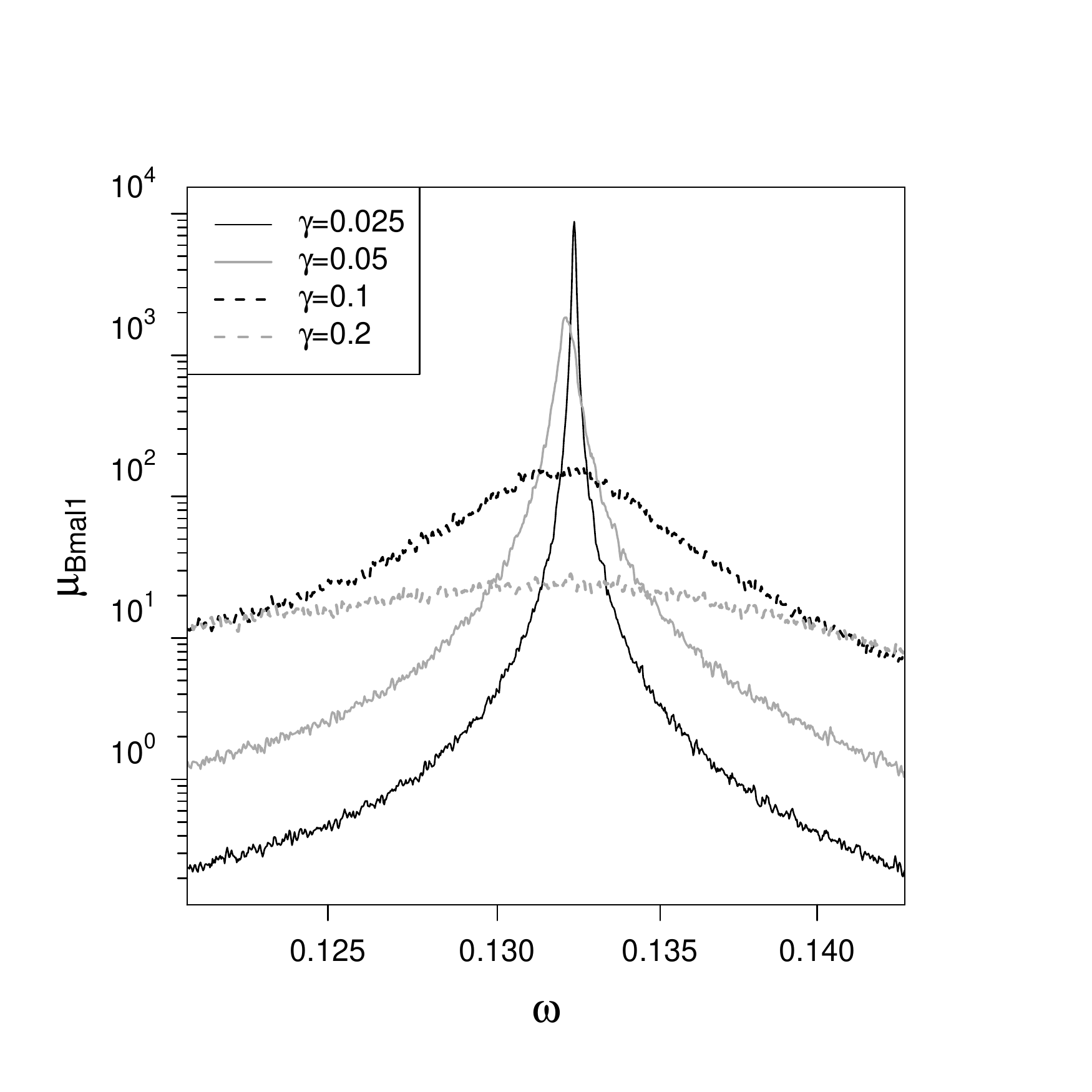}
		\caption{\label{two}Power spectral density $\mu_{Bmal1}(w)$ of the $Bmal1$ mRNA concentration trajectories, for different values of the photic perturbation intensity parameter $\gamma$.}
	\end{center}
\end{figure}

A sample realization of the dynamics with a photic perturbation fluctuating with intensity $\gamma=0.05$ is plotted in Fig.\ref{one}. While $Per2$ is directly influenced by the photic perturbation (see \eqref{Circadian Model}), $Bmal1$ is influenced only indirectly through $Per2\dashv Rev$-$erb\alpha \dashv Bmal1$ and longer paths. The continuous photic perturbation modifies the trajectories from being regular allowing oscillations to occur statistically on different periods than 24 hours. This is seen studying the spectral content of trajectories for different values of the perturbation intensity parameter $\gamma$. Let us recall the definition of power spectral density $\mu_{y}(w)$ of a process $y$ as a function of the frequency $w$:
\begin{equation}\label{PSD}
\mu_{y}(w)= \lim_{T\rightarrow \infty} \frac{\left\langle |\int_0^T dt~ e^{-i w t} y(t)|^2 \right\rangle}{T}.
\end{equation}
We see in Fig.\ref{two} that the power spectral density of variable $Bmal1(t)$, that is $\mu_{Bmal1}(w)$, has a sharp peak at around $\frac{1}{24 h}$ for small values of $\gamma$ , and that broadens when $\gamma$ is increased up to values where stable oscillations are practically lost. This effect is even larger on the light sensor $Per2$ (see Supplementary Fig.A1), indicating that the photic perturbation propagates through the circadian clock network, and is attenuated by the feedback dynamics \eqref{Circadian Model} preserving robust oscillations in the other genes.

The circadian clock model coupled to the photic perturbation dynamics \eqref{Circadian Model} is a stochastic stationary process. Let us consider the time-invariant joint probability density $p(\zeta_{\tau}^{xy})=p(x_t,y_t,x_{t+\tau},y_{t+\tau})$ of the photic perturbation $x$ and one of the circadian variables $y$, taken at two time instants separated by an interval $\tau$. We defined the combination $\zeta_{\tau}^{xy}$, and the observational time $\tau$ that specifies the stationary time series framework \cite{auconi2019information}. The system \eqref{Circadian Model} is not Markovian due to the time-delayed interactions, therefore the joint probability at two time instants $p(\zeta_{\tau}^{xy})$ cannot be a complete description of the dynamics. System \eqref{Circadian Model} could be expressed in Markovian form if we would consider portions of trajectories, for each variable of a lenght equal to its interaction time delay $\tau_i$. For a time delay of interactions that is comparable to the characteristic time of the dynamics (that is the period of oscillations), which is our case, this approach would not be computationally feasible even for a single variable. If we then consider only the two time points statistics, still the probability density of the 12-dimensional variable $\zeta_{\tau}^{x\vec{y}}$ cannot be estimated with the precision needed to compare irreversibility and information-theoretic measures. We will therefore consider $p(\zeta_{\tau}^{xy})$ for one variable $y$ at a time, and varying the observational time $\tau$ we wish to gain insight into the photic perturbation propagation through the circadian network. Note that since we consider only one of the circadian variables at a time, the conditional probability $p(y_{t+\tau}|x_t,y_t)$ has a larger variance compared to the full knowledge of the state at time $t$, $p(y_{t+\tau}|x_t,\{y_j(t)\}_{\small j=1,...,5})$.

As we already mentioned, irreversibility measures are based on time-reversal asymmetries.
Let us define the backward combination $\widetilde{\zeta_{\tau}^{xy}}$ as the time-reversal states of $\zeta_{\tau}^{xy}$, namely $\widetilde{\zeta_{\tau}^{xy}}\equiv(x_{t+\tau},y_{t+\tau},x_t,y_t)= (x(t)=x_{t+\tau},y(t)=y_{t+\tau},x(t+\tau)=x_t,y(t+\tau)=y_t)$.
The stochastic mapping irreversibility \cite{auconi2019information} for system $(x,y)$ is defined as:
\begin{eqnarray}\label{Definition}
\varphi_\tau^{xy}=\ln \left(\frac{p(\zeta_\tau^{xy})}{p(\widetilde{\zeta_{\tau}^{xy}})}\right).
\end{eqnarray}
$\varphi_\tau^{xy}$ depends on the particular realization $\zeta_{\tau}^{xy}$ and on the ensemble distribution which specifies the form of $p(\zeta_{\tau}^{xy})$. The mapping irreversibility is defined as the ensemble average of its stochastic counterpart, $\varPhi_\tau^{xy}\equiv \langle  \varphi_\tau^{xy}\rangle$. Defining $\zeta_{\tau}^{x}\equiv (x_t,x_{t+\tau})$, then the stochastic mapping irreversibility for the $x$ variable alone is simply $\varphi_\tau^{x}=\ln \left(\frac{p(\zeta_\tau^{x})}{p(\widetilde{\zeta_{\tau}^{x}})}\right)$, and for $y$ an analogous expression holds.

Importantly, even if the underlying dynamics would be Markovian bipartite, that means conditionally independent in updating $p(x_{t+dt},y_{t+dt}|x_t,y_t)=p(x_{t+dt}|x_t,y_t)\cdot p(y_{t+dt}|x_t,y_t)$, the observation at a finite resolution $\tau>0$ makes the corresponding time series non-bipartite in general, $p(x_{t+\tau},y_{t+\tau}|x_t,y_t)=p(x_{t+\tau}|x_t,y_t)\cdot p(y_{t+\tau}|x_t,y_t,x_{t+\tau})$. This is what makes the time series formulation different from the continuous stochastic thermodynamics: that probabilities cannot be expressed in terms of Onsager-Machlup action functionals \cite{rosinberg2016continuous,onsager1953fluctuations}. Let us note that in our case, the underlying dynamics \eqref{Circadian Model} is anyway not bipartite if we consider the time-delayed conditions and the absence of a noise source in the response \cite{auconi2019information}.

The circadian oscillations $y(t)$ are time-asymmetric even in the absence of perturbations ($\gamma=0$) due to the non trivial form of the $f_i$ in \eqref{Circadian Model}, and this is reflected in the $y$ mapping irreversibility being positive, $\varPhi_{\tau}^y>0$. The joint irreversibility is lower bounded by that of the subsystems, $\varPhi_{\tau}^{xy}\geq\varPhi_{\tau}^y>0$, and is therefore not the right measure to quantify the influence of photic perturbations. We wish to remove the intrinsic asymmetry of such nonlinear oscillations, and to only consider that fraction of irreversibility that results from the continuous photic perturbation. Therefore, in analogy with the definition of mutual entropy production given in  \cite{diana2014mutual}, we define the Markovian approximation to it considering only the statistics of single steps in the time series, and we call it mutual mapping irreversibility $\Theta_\tau^{xy}\equiv \langle \theta_\tau^{xy} \rangle$. Its stochastic realization-dependent counterpart is written:
\begin{equation}\label{stochastic mutual mapping irreversibility}
\theta_\tau^{xy}\equiv \varphi_\tau^{xy}-\varphi_\tau^x-\varphi_\tau^y.
\end{equation}
$\Theta_\tau^{xy}$ is the amount of mapping irreversibility in the joint time series that is due to the interaction between subsystems.

Our circadian system \eqref{Circadian Model} is a signal-response model \cite{auconi2017causal} because the dynamics of the photic perturbation $x$ is not affected by any of the circadian variables $y_i$. For signal-response models an inequality holds \cite{auconi2019information} that sets the backward transfer entropy \cite{ito2016backward} as a lower bound to the conditional mapping irreversibility, $\varPhi_\tau^{y|x}\equiv \varPhi_\tau^{xy}-\varPhi_\tau^x\geq T_{y\rightarrow x}(-\tau)$. The backward transfer entropy is defined as the standard transfer entropy for time-reversal trajectories \cite{ito2016backward}:
\begin{equation}
	T_{y\rightarrow x}(-\tau)\equiv \left\langle \ln \left( \frac{p(x_t|y_{t+\tau},x_{t+\tau})}{p(x_t|x_{t+\tau})} \right) \right\rangle.
\end{equation}
Therefore for the mutual mapping irreversibility it holds:
\begin{equation}
\Theta_\tau^{xy}\geq T_{y\rightarrow x}(-\tau)-\varPhi_\tau^y\geq -\varPhi_\tau^{xy}.
\end{equation}
This does not necessarily provide a positive lower bound to the mutual mapping irreversibility since $\varPhi_\tau^y$ is often larger than $T_{y\rightarrow x}(-\tau)$. Indeed $\Theta_\tau^{xy}$ is not defined positive \cite{diana2014mutual}, and the general lower-bound is $\Theta_\tau^{xy}\geq -\varPhi_\tau^{xy}$.

\begin{figure}
	\begin{center}
		\includegraphics[scale=0.55]{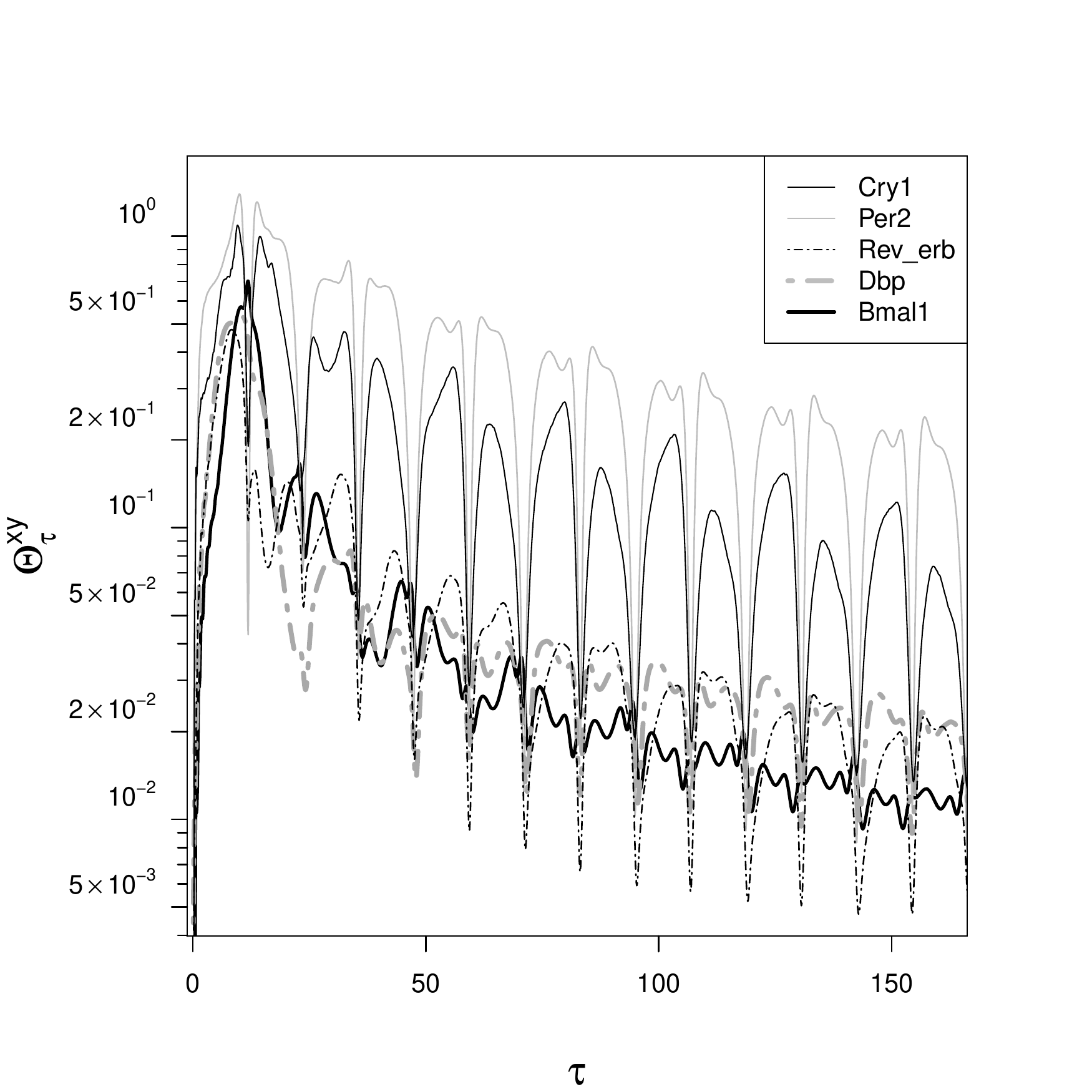}
		\caption{\label{three} Mutual mapping irreversibility $\Theta_\tau^{xy}$ for the five circadian variables as a function of the observational time $\tau$, for photic perturbation fluctuations of intensity $\gamma=0.05$.}
	\end{center}
\end{figure}

$\Theta_\tau^{xy}$ is our quantitative description of the influence of continuous photic perturbations on circadian rhythms. The explicit dependence on $\tau$ tells us how the effects are observed over time. In our numerical experiment $\Theta_\tau^{xy}$ results to be always positive for the system \eqref{Circadian Model}, $\Theta_\tau^{xy}\geq 0$.
In Fig.\ref{three} we plot $\Theta_\tau^{xy}$ for the five genes as a function of the observational time $\tau$, for perturbation fluctuations of intensity $\gamma=0.05$.

$\Theta_\tau^{xy}$ vanishes for $\tau\rightarrow 0$ because of the uncertainty in the dynamics which derives from the other four non considered variables. More explicitly, for small $\tau$ (and $\gamma>0$) the distribution $p(x_{t+\tau},y_{t+\tau}|x_t,y_t)$ is bimodal and converges (in the Kullback-Leibler sense) to $p(x_{t-\tau},y_{t-\tau}|x_t,y_t)$, while $p(x_{t+\tau},y_{t+\tau}|x_t,\vec{y_t})$ is unimodal and diverges from $p(x_{t-\tau},y_{t-\tau}|x_t,\vec{y_t})$.

$\Theta_\tau^{xy}$ increases for all variables for small $\tau$, much before the delay time of interactions with $Per2$, $\tau<\tau_{\small Per2}$, because of the correlation time of the signal. In other words, the knowledge of signal state at time $t$, that is $x_t$, gives a non-negligible amount of information on the signal at previous time instants $t-\tau\sim t-t_{rel}$, which then gives information on the other variables at previous times where their delayed influence on time $t$ matters.

Note that for the circadian model of self sustained oscillations \eqref{Circadian Model}, and especially for small perturbation intensities ($\gamma<0.1$), it results that  $\Theta_\tau^{xy}<<\varPhi_\tau^{xy}$. This means that the irreversibility due to the interaction is only a small fraction of the total irreversibility, that fraction being captured by the mutual irreversibility $\Theta_\tau^{xy}$.

We see that, after a transient period of roughly 48 hours, the mutual entropy production $\Theta_\tau^{xy}$ shows periodic regular patterns for all genes while exponentially decaying. We can factor $\Theta_\tau^{xy}$ assuming the form $\Theta_\tau^{xy}=A e^{-B\tau} f(\tau)$, where $A>0$ is the $\Theta_\tau^{xy}$ intensity, $B$ is the $\Theta_\tau^{xy}$ decay rate, and $f(\tau)$ is the $\Theta_\tau^{xy}$ oscillating component.
$Per2$ has the highest intensity $A_{\small Per2}=0.58$ being the direct sensor of photic perturbations; it is followed by $Cry1$ with $A_{\small Cry1}=0.32$, and this can possibly be related with the centrality of $Cry1$ being the only variable that is influenced by all the others. The remaining variables have a much smaller response to photic perturbations, $A_{\small Dbp}\approx A_{\small Rev\smash{-}erb\alpha}\approx A_{\small Bmal1}\approx 0.05$. The decay rates are almost equal for all variables $B_{\small Per2} \approx B_{\small Cry1} \approx B_{\small Dbp}\approx B_{\small Rev\small{-}erb\alpha}\approx B_{\small Bmal1} \approx 0.01$, meaning that in the long term any perturbation spreads its effect to all the variables while being attenuated. In order to characterize the oscillating component $f(\tau)$ we study its spectral content with the single realization discrete PSD. The PSD results to have a strong peak for the harmonics corresponding to the 12 hour period, and smaller peaks for 6, 24, 4, and 8 hours periods. We extract the characteristic period of the mutual irreversibility oscillations $T_{\Theta}$ as a weighted average of the corresponding harmonics, $T_{\Theta}=\frac{1}{N_{PSD}}\sum_{j=1}^{\infty}\frac{PSD(i)}{ w(i)}$ with normalization factor $N_{PSD}=\sum_{j=1}^{\infty}PSD(i)$. The characteristic period of oscillations is around 9-13 hours for all the variables.
The result $T_{\Theta}\approx 12 h$ means that, while the dynamics is strongly characterized by 24 hours oscillations, the response to perturbations is dominated by 12 hour harmonics. This is related to a previous result \cite{westermark2013mechanism}, and is understood considering the form of the mapping irreversibility \eqref{Definition} with the $24~h$ periodicity $y_t\approx y_{t+T}$, for which we can write the symmetry at $\tau=\frac{T}{2}$: $p(y_t,y_{\frac{T}{2}})\approx p(y_t,y(t-\frac{T}{2})=y_{\frac{T}{2}})= p(\widetilde{y_t},\widetilde{y_{\frac{T}{2}}})$, and therefore $\varPhi_{\tau}^y$ vanishes twice every 24h period. This structure is preserved in $\varPhi_{\tau}^{y|x}$ and in $\Theta_\tau^{xy}=\varPhi_{\tau}^{y|x}- \varPhi_{\tau}^y$, meaning that the dynamics of the integrated response to perturbations maintains the oscillatory property of the unperturbed dynamics.

If we modify the relaxation time of fluctuations $t_{rel}$ keeping the same intensity and standard deviation for $x$, the mapping irreversibility structure is qualitatively preserved, with the response amplitudes $A$s increasing with $t_{rel}$ (see Supplementary Fig.A2-A3). The structure is also preserved varying $\gamma$, with the decay rates $B$s increasing with $\gamma$, consistent with the loss of rhythmicity observed in the dynamics (see Supplementary Fig.A4-A5). Importantly, while $\varPhi_{\tau}^{xy}$ decreases with $\gamma$, the first peak in the mutual mapping irreversibility $\Theta_\tau^{xy}$  is not.

Let us mention that the response to perturbations in oscillating systems has been considered for network reconstruction \cite{timme2007revealing}. The response to fluctuating signals might be even more suited for the task, but that is beyond the scope of this paper.

\begin{figure}
	\begin{center}
		\includegraphics[scale=0.55]{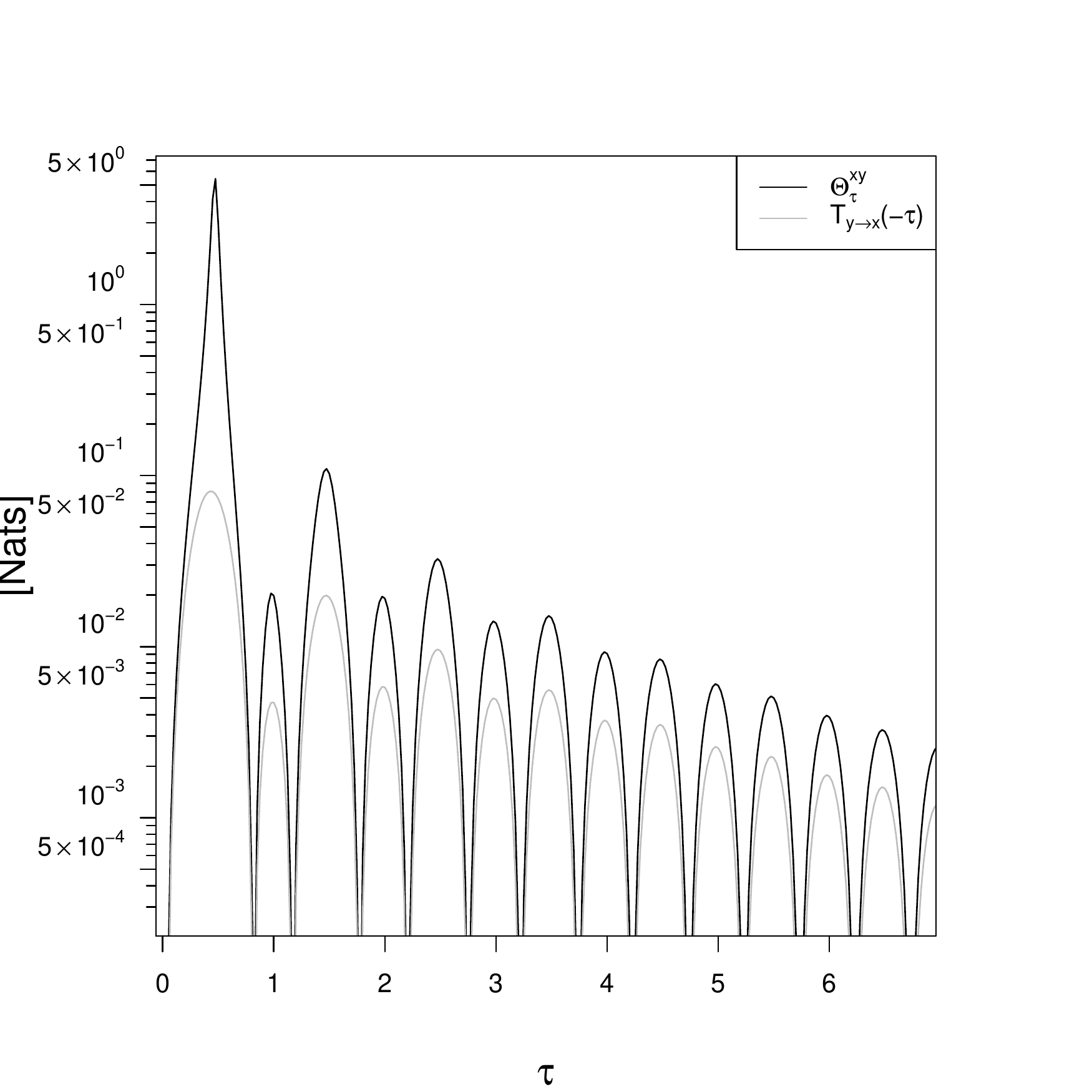}
		\caption{\label{four} Mutual mapping irreversibility $\Theta_\tau^{xy}$ in the damped linear oscillator driven by colored noise \eqref{dumped oscillator}, with parameters $t_{rel}=1$, $\beta=0.2$, and $\gamma=1$. In gray we plot the backward transfer entropy $T_{y\rightarrow x}(\tau)$, that is the lower bound given by the time series fluctuation theorem \cite{auconi2019information}.}
	\end{center}
\end{figure}

Let us show here that the 12 hour harmonics in the mutual irreversibility is not due to the nonlinear behavior, and also not on the self-sustained property. Indeed, they are observed also in a linear damped oscillator $y$ driven by colored noise $x$:
\begin{eqnarray}\label{dumped oscillator}
	\begin{cases} 
		dx =-\frac{x}{t_{rel}} dt + dW \\
		\frac{dy}{dt} = -\beta y + \gamma x ~-(2\pi)^2 \int_{-\infty}^{t} dt'~ y_{t'} e^{-\beta (t-t')} 
	\end{cases}
\end{eqnarray}
where $\beta>0$ and the term $(2\pi)^2$ sets the oscillations' period to $1$. Here the mutual irreversibility peaks occur every half period ($\frac{1}{2}$ units), see Fig.\ref{four}, and that is also the case for mutual information and transfer entropy measures. A correspondence can be suggested with the theory of attractor embedding in chaotic deterministic systems \cite{fraser1986independent,packard1980geometry,vulpiani2010chaos}, where delays at half the period and multiples correspond to poor projections on the attractor \cite{gober1992dimension}.

In model \eqref{dumped oscillator} both the subsystem's dynamics is time symmetric, $\varPhi_{\tau}^x=0$ and $\varPhi_{\tau}^y=0$, and the irreversibility is seen in the interaction and found in the joint mapping irreversibility $\varPhi_{\tau}^{xy}$. Then $\Theta_{\tau}^{xy}=\varPhi_{\tau}^{xy}$ and the inequality with the backward transfer entropy reads $\Theta_{\tau}^{xy}\geq T_{y\rightarrow x}(\tau)$, and it is plotted in Fig.\ref{four}. Let us also note that, similar to what we found in the circadian clock model (Fig.\ref{three}), the asymmetry of successive peaks decreases with time. The difference in the response to fluctuations between the nonlinear circadian model \eqref{Circadian Model} and the linear damped oscillator \eqref{dumped oscillator} is in the shapes of curves, that look indeed non trivial in the circadian clock mutual irreversibility (Fig.\ref{three}).


To summarize, we applied the time series stochastic thermodynamics framework to quantify the influence of photic perturbations on circadian rhythms.
In particular, we considered the main effect of such asymmetric (causal) interactions, that is the irreversibility of time series, over different time scales $\tau$. It is captured by the mutual mapping irreversibility measure \eqref{stochastic mutual mapping irreversibility}, and results to be characterized by half period harmonics. Its magnitude on the different circadian genes is consistent with the network topology, with $Cry1$ being the most influenced gene after direct photic perturbations on $Per2$.

\textbf{Acknowledgements}
We thank M Scazzocchio for useful discussions.
Work at Humboldt-Universit\"at zu Berlin was supported by the DFG (Graduiertenkolleg 1772 for Computational Systems Biology).

\bibliography{clock}

\end{document}